\documentstyle[prl,twocolumn,aps,floats]{revtex}

\begin{document}

\draft

\title{Spreading Width for Decay out of a Superdeformed Band}
\author{H. A. Weidenm\"uller$^{1}$, P. von Brentano$^{2}$, and
  B. R. Barrett$^{3}$}
\address{$^{1}$Max-Planck-Institut f\"ur Kernphysik, 69028 Heidelberg,
  Germany}
\address{$^{2}$Institut f\"ur Kernphysik, Universit\"at zu K\"oln, 50937 K\"oln,
Germany}
\address{$^{3}$Department of Physics, University of Arizona, Tucson, AZ 85721
USA.}

\date{\today}

\maketitle

\begin{abstract}
The attenuation factor $F$ responsible for the decay out of a
superdeformed (SD) band is calculated with the help of a statistical
model. This factor is given by $F = (1 + \Gamma^{\downarrow} / 
\Gamma_{\rm S})^{-1}$. Here, $\Gamma_{\rm S}$ is the width for the
collective E2 transition within the superdeformed band, and
$\Gamma^{\downarrow}$ is the spreading width which describes the
mixing between a state in the SD band and the normally deformed (ND)
states of equal spin. The attenuation factor $F$ is independent of the
statistical E1 decay widths $\Gamma_{\rm N}$ of the ND states provided
that $\Gamma_{\rm N} \gg \Gamma^{\downarrow}, \Gamma_{\rm S}$. This
condition is generically met. Previously measured values of $F$ are
used to determine $\Gamma^{\downarrow}$.
\end{abstract}

\medskip

The intensities of E2 gamma transitions within a superdeformed (SD)
rotational band show a remarkable feature. The intraband E2 transitions
follow the band down with practically constant intensity. At some
point, the transition intensity starts to drop sharply. This
phenomenon is referred to as the decay out of a superdeformed
rotational band. It is attributed to a mixing of the SD states and the
normally deformed (ND) states with equal spin. The barrier separating
the first and second minima depends on and decreases with
decreasing spin I. Decay out of the SD band sets in at a spin value
I$_0$ for which penetration through the barrier is competitive with
the E2 decay within the SD band. The theoretical description of this
process \cite{sch89,vig90,vig90a,shi92,shi93,kho93} uses a statistical
model for the ND states. The actual physical decay out of the
superdeformed band is calculated as a function of the parameters of
the model. Recent experimental data given in
Refs.~\cite{kru94,kru96,kuh97,kru97} have been analyzed using some of
these results.

In the present paper, we take a different approach: We calculate the
reduction factor $F$ of the intraband transition intensity (an
observable) directly in terms of the spreading width
$\Gamma^{\downarrow}$ and of the intraband E2 width $\Gamma_{\rm S}$
for electromagnetic decay of the SD state. The spreading width
$\Gamma^{\downarrow}$ measures the strength of the coupling between
the SD state and the ND states. We show that $F$ does not depend on
the electromagnetic decay properties of the ND states, provided that
the width for statistical E1 decay of the ND states is much larger
than $\Gamma_{\rm S}$. Therefore, a measurement of $F$ directly yields
the spreading width $\Gamma^{\downarrow}$ and, thus, direct
information on properties of the barrier separating the first and 
second minima. This information is practically parameter--free and
model--independent. It is obtained without going through a complete
model calculation of the entire process leading to decay out of a SD
rotational band. We use data given in
Refs.~\cite{kru94,kru96,kuh97,kru97} to deduce values for 
$\Gamma^{\downarrow}$ in $^{192,194}$Hg and $^{194}$Pb. We test our
procedure by comparing the values of $\Gamma_{\rm S}$ obtained in that
same analysis with predictions based on properties of the SD band. Our
approach and our results differ from those of
Refs.~\cite{vig90,vig90a,shi92,shi93}. We comment on these differences
at the end of the paper.

To define the model, we denote the first SD state with significant
coupling to the ND states during the E2 decay down the
SD band by $|0 \rangle$; its energy by $E_0$; the ND states having the
same spin as the state $|0 \rangle$ by $|j \rangle$ with $j = 1,
\ldots,K$; their energies by $E_j$; and the average spacing of the ND
states by $D$. We take the limit $K \rightarrow \infty$ in the course
of the calculation. The ND states decay by statistical E1 emission. We
assume that the total E1 decay widths of all ND states have identical
values denoted by $\Gamma_{\rm N}$. The coupling matrix elements
$V_{0j}$ connect the SD and the ND states and are responsible for
decay out of the SD band. Following
Refs. \cite{vig90,vig90a,shi92,shi93}, we assume that the ND states
$|j\rangle$ can be modeled as eigenstates of the Gaussian Orthogonal
Ensemble (GOE) of random matrices. The spreading width
$\Gamma^{\downarrow}$ is defined as $\Gamma^{\downarrow} = 2 \pi v^2 /
D$, where $v^2$ is the mean square of the matrix elements $V_{0j}$. 
Typically, we have $\Gamma_{\rm N}, \Gamma_{\rm S} \ll D$, see Table
I. We note that in all cases, $\Gamma_{\rm S} \ll \Gamma_{\rm
  N}$. Because of the dominance of E1 over all other transitions, this
inequality is expected to hold generically. It is fair to expect that
decay out of the SD band will set in whenever decay time
$\hbar/\Gamma_{\rm S}$ and mixing time $\hbar/\Gamma^{\downarrow}$ of
the state $|0\rangle$ are comparable, i.e., when $\Gamma^{\downarrow}$
is of the same order of magnitude as $\Gamma_{\rm S}$. Thus, we expect
that we also have $\Gamma^{\downarrow} \ll D$. This is, indeed, borne
out by the analysis described below.

The Hamiltonian $H$ of the system is modeled as a matrix of dimension
$K + 1$ and has the form  
\begin{eqnarray} 
&&H\!=\!\left(\!
\begin{array}{ll}
E_0&V_{0 j}\\
V_{0 l}&\delta_{j l} E_j\end{array}
\!\right)\! \ .
\end{eqnarray}
To $H$ must be added the diagonal width matrix $\Sigma$ with
\begin{eqnarray} 
&&\Sigma\!=\!-(i/2)\left(\!
\begin{array}{ll}
\Gamma_{\rm S}&0\\
0&\delta_{j l} \Gamma_{\rm N}\end{array}
\!\right)\! \ .
\end{eqnarray}
The effective Hamiltonian ${\cal H}$ is given by ${\cal H} = H +
\Sigma$. 

We first investigate an experimental situation which would require an
energy resolution that is not available at present: We study the
distribution in energy of the E2 transition intensity {\it feeding}
the state $|0\rangle$. Since $\Gamma^{\downarrow} \neq 0$, this state
is mixed with the ND states, and the E2 transition intensity feeding
the state is spread out over a number of eigenstates of ${\cal H}$. We
calculate the ensemble average of this intensity distribution. The
distribution cannot be measured at present because the required
resolution is of the order of the average spacing $D$ of the ND
states. However, the calculation defines concepts and yields results
which are important later on. 

We first consider the case $\Gamma_{\rm N} = 0$. For
$\Gamma^{\downarrow} \ll D$, the mixing of the SD state with the ND
states will cause the E2 feeding strength to have a central peak
located at (or near) $E_0$, and a number of much smaller and
well--separated peaks located at (or near) the energies $E_j$. The 
central peak will obviously be reduced in height compared to the case 
$\Gamma^{\downarrow} = 0$. If observable, this phenomenon could be
referred to as ``decay out of the SD band'': The peak E2 transition
intensity is reduced. The reduction occurs even when the ND states are
not capable of decaying by statistical E1 emission. If we allow for
the statistical E1 decay of the ND states, the picture remains
essentially the same except that all peaks become wider. We expect
that the reduction of the E2 transition peak intensity is not
dependent on $\Gamma_{\rm N}$ and is governed by the competition
between mixing time and decay time, i.e., by the ratio
$\Gamma^{\downarrow}/\Gamma_{\rm S}$. We show presently that this is
indeed the case.

To calculate the distribution of transition strength, we again consider
first the case $\Gamma^{\downarrow} = 0$. The state $|0\rangle$ is fed
from the next higher SD state. The relative transition intensity
$I(E)$ versus energy $E$ is given by 
\begin{equation}
I(E) = - 2 \gamma \ {\sl [Im}\ (E - E_0 + (i/2) \Gamma_{\rm S} )^{-1}] \
\gamma
\label{intensity}
\end{equation}
where ${\sl Im}$ (or, later, ${\sl Re}$) stands for the imaginary (the
real) part. The symbol $\gamma$ denotes the E2 decay amplitude
feeding the state $|0\rangle$, and $I(E)$ has the expected Lorentzian
form. Integrating over all $E$, we obtain $\int dE I(E) = 2 \pi
\gamma^2$. We extend Eq.~(\ref{intensity}) to the case
$\Gamma^{\downarrow} \neq 0$ by writing it as 
\begin{equation}
I(E) = - 2 \gamma \ [{\sl Im}\ ({\bf E} - {\cal H})^{-1}]_{00} \ \gamma
\ , 
\label{intensity1}
\end{equation}
where ${\bf E} = E \times 1_{K+1}$ and where $1_K$ is the $K$--dimensional
unit matrix. To check the general validity of Eq.~(\ref{intensity1}),
we transform ${\cal H}$ to diagonal form with left (right)
eigenfunctions $\psi_m^L (\psi_m^R)$ and complex eigenvalues ${\cal
  E}_m$ with $m = 0,\ldots,K$. Then, Eq.~(\ref{intensity1}) takes the form
\begin{equation}
I(E) = - 2 \gamma \sum_{m=0}^K {\sl Im} \ [ \langle 0|\psi_m^R \rangle \
  (E - {\cal E}_m)^{-1} \langle \psi_m^L| 0 \rangle ] \ \gamma \ , 
\label{intensity2}
\end{equation}
which is obviously a sum of Lorentzians. The $m^{\rm th}$ Lorentzian
is peaked at the energy $E = {\sl Re} \ {\cal E}_m$. The area under
the peak is given by $2 \pi$ times the probability $\langle 0 |
\psi_m^R \rangle \langle \psi_m^L | 0 \rangle $ to find the state
$|0\rangle$ admixed into the $m^{\rm th}$ eigenstate. A change of the
E1 width $\Gamma_N$ will result in a change of the peak widths of the
$m$ resonances and the weight factors $\langle 0 |\psi_m^R \rangle
\langle \psi_m^L | 0 \rangle$. The total intensity (the sum of the
contributions of all $m$ resonances) is given by $2 \pi \gamma^2$, as
in the case $\Gamma^{\downarrow} = 0$. It is independent of
$\Gamma_N$. These results show that we are dealing with a spreading
width phenomenon, which is not related to the E1 decay of the ND
states.

It remains to average $I(E)$ over the GOE. We recall that the energies
$E_j$ are the eigenvalues of a GOE matrix, and that, correspondingly,
the matrix elements $V_{0j}$ are uncorrelated Gaussian distributed
random variables with common mean value zero and common variance $v^2$. 
We denote the ensemble average by a bar. Clearly, all resonance
structure will be washed out in $\overline{I(E)}$, and physical
intuition and the arguments given above lead us to expect that
$\overline{I(E)}$ has Lorentzian shape and width $\Gamma_S +
\Gamma^{\downarrow}$. This is, indeed, the result of the calculation. 
To see this, we put first $E = E_0$ and write
\begin{equation}
\label{propagator}
[({\bf E}_0 - {\cal H})^{-1}]_{00} = -(i/\Gamma_S) \ [1 + S_{00}(E_0)]
\ .
\end{equation}
Here, $S_{00}(E)$ is given by
\begin{equation}
S_{00}(E) = 1 - 2 \pi i \sum_{j,l} w_{0j} [D^{-1}(E)]_{jl} w_{l0}
\label{smatrix}
\end{equation}
where $w_{0j} = w_{j0} = \sqrt{2 / (\pi \Gamma_{\rm S})} \ V_{0j}$ and
\begin{equation}
\label{prop}
D_{jl}(E) = (E - E_j + (i/2)\Gamma_{\rm N})\delta_{jl} + i \pi w_{j0}
w_{0l} \ .
\end{equation}
Equations~(\ref{smatrix},\ref{prop}) show that $S_{00}$ is an element
of a bona fide unitary scattering matrix. Aside from the channel
denoted by zero, there are inelastic channels corresponding to the
emission of E1 radiation by the ND states and leading to the
appearance of the term $\Gamma_{\rm N}$ in Eq.~(\ref{prop}). We have
introduced the form Eq.~(\ref{propagator}) because calculating the
ensemble average over the diagonal element of a scattering matrix
defined in terms of the GOE is a standard problem in stochastic
scattering theory. The ensemble average is taken \cite{ver85} over the
distribution of the $V_{0j}$'s and $E_j$'s and yields
\begin{equation}
\overline{S_{00}(E)} = \frac{E - E_0 + (i/2)(\Gamma_{\rm S} -
  \Gamma^{\downarrow})}{E - E_0 + (i/2)(\Gamma_{\rm S} +
  \Gamma^{\downarrow})} \ .
\label{saverage}
\end{equation}
For the reader not familiar with the technicalities of ref.~\cite{ver85}, 
we mention that Eq.~(\ref{saverage}) can be obtained in a simpler way: We 
replace the ensemble average by the substitution $E - E_j \rightarrow 
E - E_j + iI$ with $I \gg D$. This is patterned after Brown's energy 
averaging procedure \cite{bro59}. However, we keep $E_0$ fixed because the 
GOE average does not affect the SD state. The independence of 
$\overline{S_{00}}$ of $\Gamma_{\rm N}$ is caused by the GOE average which 
replaces the discrete spectrum of the ND states by a quasi--continuum. 
Using this result for averaging Eqs.~(6) and (4), we find
\begin{equation}
\label{result}
\overline{I(E)} = - 2 \gamma \ {\sl [Im} \ (E - E_0 + (i/2)
(\Gamma_{\rm S} + \Gamma^{\downarrow}))^{-1}] \ \gamma \ .
\end{equation}
Equation~(\ref{result}) shows that the average intensity has
Lorentzian shape with width $\Gamma_{\rm S} + \Gamma^{\downarrow}$ and
is independent of $\Gamma_{\rm N}$. The ratio of $\overline{I(E_0)}$
and the intensity for $\Gamma^{\downarrow} = 0$ at $E = E_0$ given by
Eq.~(3) yields the average peak intensity attenuation factor $F$,
\begin{equation}
\label{factor}
F = (1 + (\Gamma^{\downarrow} / \Gamma_{\rm S}))^{-1} \ .
\end{equation}

The peak attenuation factor $F$ does possess physical significance but
cannot be measured in practice. However, $F$ also determines the
reduction of the total intensity of E2 radiation down from the state
$|0\rangle$ and the states mixed with it into the next state of the SD
band. This information is presently available and used below to
determine $F$ from the data. We have just shown that it is not
necessary to ask in which way the next lower SD state is mixed with
the ND states when we ask for the total intensity feeding it. It
suffices to calculate the transition intensity feeding that next lower
SD state.

In the simplest case where $\Gamma^{\downarrow} = 0$, only the
intermediate population of the state $|0\rangle$ is of interest. 
With $\Gamma_{\rm S} = \gamma_{\rm S}^2$, the transition amplitude has
the form 
\begin{equation}
T(E) = \gamma \ (E - E_0 + (i/2) \Gamma_{\rm S})^{-1} \ \gamma_{\rm S}
\ .
\end{equation}
We have $\int dE |T(E)|^2 = 2 \pi \gamma^2$: Feeding intensity and
emitted intensity are, of, course, equal. For $\Gamma^{\downarrow}
\neq 0$, we have 
\begin{equation}
\label{transition}
T(E) = \gamma \ [({\bf E} - {\cal H})^{-1})]_{00} \ \gamma_{\rm S} \ .
\end{equation}
For $\Gamma_{\rm N} = 0$, the total strength $\int \ dE \ |T(E)|^2 = 2
\pi \gamma^2$ is expected to be preserved. This is, indeed, the case
and can be checked by using the identity
\begin{eqnarray}
\label{identity}
[({\bf E} - {\cal H})^{-1})]_{00} \Gamma_{\rm S} [({\bf E} - {\cal
  H^*})^{-1})]_{00} = \nonumber \\
- 2 \ {\sl Im} \ [({\bf E} - {\cal H})^{-1})]_{00} \ . 
\end{eqnarray}
The identity Eq.~(\ref{identity}), in turn, follows immediately when
we use the form $[({\bf E} - {\cal   H})^{-1})]_{00} = (E - E_0
  +(i/2)\Gamma_{\rm S} - \sum_j V_{0j} (E - E_j)^{-1}
  V_{j0})^{-1}$. Averaging over the ensemble does not affect the
conservation of total strength. A change does occur, however, for
$\Gamma_{\rm N} \neq 0$. Then, the identity Eq.~(\ref{identity}) does
not apply and E1 decay of the ND states weakens the E2 decay
intensity. The question is: By how much?

To calculate $\int \ dE \ \overline{|T(E)|^2}$ for $\Gamma_{\rm N}
  \neq 0$, we use the form $[({\bf E} - {\cal H})^{-1}]_{00} = (1/2)
  (E - E_0 +(i/2) \Gamma_{\rm S} )^{-1} \ [ 1 + S_{00}(E) ]$, see
Eqs.~(6,7), and write $S_{00} = \overline{S_{00}} + S_{00}^{\rm fl}$
where the upper index fl denotes the fluctuating part. Then, 
\begin{eqnarray}
\label{average}
&&\overline{|T(E)|^2} = \gamma^2 \Gamma_S \times \nonumber \\
&&(1/4) [(E - E_0)^2 + (1/4) \Gamma_{\rm S}^2]^{-1} (|1 +
\overline{S_{00}}|^2 + \overline{|S_{00}^{fl}|^2}) \ .
\end{eqnarray}
The entities on the rhs of Eq.~(\ref{average}) are all known:
$\overline{S_{00}}$ is given in Eq.~(\ref{saverage}), and
$\overline{|S_{00}^{fl}|^2}$ is given explicitly in Ref.~\cite{ver85}. 
Integration over $\overline{|T(E)|^2}$ then yields the E2 intensity
down to the next SD state. The actual calculation would be rather
involved. It would involve a fourfold numerical integration. 
Fortunately, the explicit calculation is not needed since for
$\Gamma_{\rm N} \gg \Gamma_{\rm S}, \Gamma^{\downarrow}$, the term
$\overline{|S_{00}^{fl}|^2}$ can be neglected. This is seen as
follows. As in any stochastic reaction problem, the fluctuating part
of the scattering matrix describes those processes where the long--lived
intermediate resonances (the ND states) undergo statistical decay. For
$\Gamma_{\rm N} \gg \Gamma_{\rm S}, \Gamma^{\downarrow}$, such decay will
overwhelmingly lead to E1 emission. The term
$\overline{|S_{00}^{fl}|^2}$ describes the statistical decay of the ND
states back into the SD state and can, therefore, be neglected. We
quantify this statement by comparing the sum $t_{\rm N}$ of the
transmission coefficients for E1 decay with the transmission
coefficient $t_0 = 1 - | \overline{S_{00}}|^2$ for decay back into the
state $|0\rangle$. For $\Gamma_{\rm N} \ll D$, we have $t_{\rm N} =
\Gamma_{\rm N} / D$, while $t_0$ depends upon energy and is given by
$t_0 = \Gamma_{\rm S} \Gamma^{\downarrow} / [(E - E_0)^2 + (1/4)
(\Gamma_{\rm S} + \Gamma^{\downarrow})^2]$. For ND resonances a
distance $D$ from $E_0$ we have $t_0 \sim  \Gamma_{\rm S}
\Gamma^{\downarrow} / D^2 \ll t_{\rm N}$. The inequality $t_0 \ll
t_{\rm N}$ remains valid down to distances $\sim 10 \Gamma_{\rm S} \ll
D$.

Neglecting $|S_{00}^{\rm fl}|^2$, using in Eq.~(\ref{average}) for
$\overline{S_{00}}$ the value of Eq.~(\ref{saverage}), and integrating
over all $E$, we obtain
\begin{equation}
\label{result1}
\int dE \overline{|T(E)|^2} =2 \pi \gamma^2 F \ ,
\end{equation} 
with $F$ given by Eq.~(\ref{factor}). This is our central result: The
intensity attenuation factor due to decay out of the SD band is given
by $F$. 

Equation~(\ref{result1}) has a simple interpretation. Decay out of the 
SD band is a sequential process with two intrinsic time scales, the
spreading time $\hbar/\Gamma^{\downarrow}$ for populating the ND
states from the SD state $|0\rangle$, and the time $\hbar/\Gamma_{\rm
  N}$ for E1 emission. The larger of these two times defines the
relevant overall time scale. This is $\hbar/\Gamma^{\downarrow}$. 
Thus, the spreading width $\Gamma^{\downarrow}$ signifies the
effective partial decay width for the total E1 decay out of the SD
band. The branching ratio for E2 decay is given by $F = \Gamma_{\rm
  S}/(\Gamma_{\rm S} + \Gamma^{\downarrow})$. The branching ratio for
the total E1 decay is correspondingly given by $(1 -F)$. This can be
verified by extending Eq.~(\ref{identity}) to the case $\Gamma_{\rm N}
\neq 0$. These facts make it possible to determine $\Gamma^{\downarrow}$
directly from the intensity attenuation within the SD band. We expect
$\Gamma^{\downarrow}$ to increase strongly as we move down the SD
band. We point out that in contrast to the problem studied in the
first half of this paper (where $F$ contains the spreading width
$\Gamma^{\downarrow}$ of the state {\it fed} by E2 radiation), $F$ now
contains the spreading width $\Gamma^{\downarrow}$ of the state from
which the E2 radiation is emitted.

For the population of the ND states $|j\rangle$ from the SD band, the
state $|0\rangle$ acts like a doorway state. Because of the barrier
separating the first and second minima, this doorway state has
similarity to an isobaric analog resonance, where isospin conservation
has the same function as the barrier. The two cases differ in that the
width $\Gamma_{\rm S}$ of the doorway state $|0\rangle$ is {\it small}
compared to the widths $\Gamma_{\rm N}$ of the ND states. For isobaric
analogue resonances, the converse situation holds. A situation similar
to the one studied above is met in vinydelene, a molecule with a shape
isomer which is analogous to the SD state \cite{lev98}. Here, however,
the inequality $\Gamma^{\downarrow} \gg \Gamma_{\rm N}$ applies.

To sum up: Our result, Eq.~(\ref{factor}), is obtained by calculating
$\overline{S_{00}}$ and by showing that for $\Gamma_N \gg
\Gamma_S,\Gamma^{\downarrow}$, $S_{00}^{\rm fl}$ is
negligible. We have applied this result to data given in
Refs.~\cite{kru94,kru96,kuh97,kru97}. The measured quantities are
$F_{\rm exp}$, the intensity reduction for E2 decay from the state
$|0\rangle$ with spin I$_0$, and the lifetime $\tau$ of that
state. We equate $F_{\rm exp}$ with the ensemble--averaged intensity
reduction factor $F$ calculated above, $F_{\rm exp} = F$. With $1 - F
= \Gamma^{\downarrow} / (\Gamma_{\rm S} + \Gamma^{\downarrow})$ and
$(\Gamma_{\rm S} + \Gamma^{\downarrow}) = \hbar / \tau$, we have
\begin{equation}
\label{width}
\Gamma^{\downarrow} = \hbar (1 - F_{\rm exp}) / \tau. 
\end{equation}
The decay width $\Gamma_{\rm S}$ is given by $\Gamma_{\rm S} = \hbar
F_{\rm exp} / \tau$. These formulae entail an error due to statistical
fluctuations of $F_{\rm exp}$. The fluctuations are caused by the
fact that in any given nucleus, the level spectrum is discrete. A very
conservative estimate of this error (based on the influence of the
closest--lying ND states) leads to an uncertainty of a factor $2$ or
$3$ for $\Gamma^{\downarrow}$. The results are shown in Table I. We
note that measurable decay out of the SD rotational band sets in
whenever $\Gamma^{\downarrow} \geq 0.1 \ \Gamma_{\rm S}$ or so. The
values of $\Gamma_{\rm S}$ are consistent with approximately constant
quadrupole moments in the SD bands. Our results support the connection
postulated in Refs.~\cite{sch89,kru94} between $\Gamma^{\downarrow}$
and the barrier separating the first and the second minima.

In Refs.~\cite{vig90,vig90a,shi92,shi93} the decay by E1 (E2) emission
is calculated by multiplying the squares of the projections of the
eigenfunctions of the Hamiltonian Eq.~(1) onto the ND (the SD) states
with $\Gamma_{\rm N} \ (\Gamma_{\rm S}$, respectively). This scheme is
expected to apply in the regime $\Gamma^{\downarrow} \gg \Gamma_{\rm
  N}, \Gamma_{\rm S}$. Motivated by the later results of
Refs.~\cite{kru94,kru96,kuh97,kru97}, we have investigated the regime
$\Gamma_{\rm N} \gg \Gamma^{\downarrow}, \Gamma_{\rm S}$. Naturally,
decay out of the SD band follows different rules in the two regimes. 
Our analysis shows that the regime $\Gamma_{\rm N} \gg
\Gamma^{\downarrow}, \Gamma_{\rm S}$ actually applies to the data 
summarized in Refs.~\cite{kru94,kru96,kuh97,kru97}.

In summary, we have shown that data on the attenuation of the
transition strength in a SD rotational band yield direct information
on the spreading width $\Gamma^{\downarrow}$ and, thus, on properties
of the barrier separating the first and second minima. This
information is practically model--independent and parameter--free.

One of us (BRB) would like to thank Hans Weidenm\"uller and the
Max--Planck--Institut f\"ur Kernphysik, Heidelberg, and Peter von
Brentano and the Institut f\"ur Kernphysik, K\"oln, for their
hospitality and partial support. He also acknowledges partial support
by NSF grant No. PHY96-05192.  One of us (PvB) wants to acknowledge
in--depth discussions on the decay of the superdeformed states with
A. Dewald and R. Kr\"ucken and partial support by the BMBF under contract
No. 060K668.

\begin{table}[htb]
\begin{center}
\leavevmode
\begin{tabular}{|c||r|r|r|c|}
Nucleus        & $\Gamma_{\rm N}$ & $D$  & $\Gamma_{\rm S}$ &
$\Gamma^{\downarrow}$ \\ \hline 
$^{192}$Hg(12) &   103            &   34 & 1.16             &
0.18       \\ 
$^{192}$Hg(10) &   103            &   30 & 0.54             &
5.44$^{\dagger}$  \\
$^{194}$Hg(12) &   181            &   92 & 1.44             &
0.97     \\ 
$^{194}$Hg(10) &   184            &   79 & $\geq$0.47       &
$\geq$ 8.9$^{\dagger}$   \\ 
$^{194}$Pb(10) &    16            & 1699 & 0.66             &
0.11       \\
$^{194}$Pb( 8) &    17            & 1549 & 0.28             &
0.09$^{\dagger}$       \\
\end{tabular}
\protect\caption{\protect\small
The spreading widths $\Gamma^{\downarrow}$ deduced from the data
reviewed in Ref.~\protect\cite{kuh97} for a number of nuclei. The spin
values of the decaying states are given in brackets. The units are
eV for $D$, and $10^{-4}$ eV for $\Gamma_{\rm N}$, $\Gamma_{\rm S}$
and $\Gamma^{\downarrow}$. The results indicated with ${\dagger}$ were
calculated with estimated lifetimes in Ref.~\protect\cite{kuh97}. The
total width $\Gamma = \Gamma_{\rm S} + \Gamma^{\downarrow}$.}
\label{tabstab}
\end{center}
\end{table}


\begin{references}

\bibitem{sch89}K. Schiffer, B. Herskind, and J. Gascon, Z. Phys. {\bf
    A 332} (1989) 17.

\bibitem{vig90}E. Vigezzi, R. A. Broglia, and T. Dossing,
  Phys. Lett. {\bf B 249} (1990) 163.

\bibitem{vig90a}E. Vigezzi, R. A. Broglia, and T. Dossing,
  Nucl. Phys. {\bf A 520} (1990) 179c.

\bibitem{shi92}Y. R. Shimizu, F. Barranco, R. A. Broglia, T. Dossing,
  and E. Vigezzi, Phys. Lett. {\bf B 274} (1992) 253. 

\bibitem{shi93}Y. R. Shimizu, E. Vigezzi, T. Dossing, and
  R. A. Broglia, Nucl. Phys. {\bf A 557} (1993) 99c.

\bibitem{kho93}T. L. Khoo, T. Lauritsen, T. Ahmad, M. P. Carpenter, P.
  B. Fernandez, R. V. F. Janssens, E. F. Moore, F. L. H. Wolfs,
  Ph. Benet, P. J. Daly, K. B. Beard, U. Garg, D. Ye, M. W. Drigert,
  Nucl. Phys. {\bf A 557} (1993) 83c.

\bibitem{kru94}R. Kr\"ucken, A. Dewald, P. Sala, C. Meier, H. Tiesler,
  J. Altmann, K. O. Zell, P. von Brentano, D. Bazzacco,
  C. Rossi-Alvarez, R. Burch, R. Menegazzo, G. de Angelis, G. Maron,
  and M. de Poli, Phys. Rev. Lett. {\bf 73} (1994) 3359.

\bibitem{kru96}R. Kr\"ucken, A. Dewald, P. von Brentano, D. Bazzacco,
  and C. Rossi-Alvarez, Phys. Rev. C {\bf 54} (1996) 1182.

\bibitem{kuh97}R. K\"uhn, A. Dewald, R. Kr\"ucken, C. Meier,
  R. Peusquens, H. Tiesler, O. Vogel, S. Kaseman, P. von Brentano,
  D. Bazzacco, C. Rossi-Alvarez, S. Lunardi, and J. de Boer,
  Phys. Rev. C {\bf 55} (1997) R1002.

\bibitem{kru97}R. Kr\"ucken, S. J. Asztalos, J. A. Becker, B. Busse,
  R. M. Clark, M. A. Deleplanque, A. Dewald, R. M. Diamond, P. Fallon,
  K. Hauschild, I. Y. Lee, A. O. Macchiavelli, R. W. MacLeod,
  R. Peusquens, G. J. Schmid, F. S. Stephens, K. Vetter, and P. von
  Brentano, Phys. Rev. C {\bf 55} (1997) R1625.

\bibitem{ver85}J. J. M. Verbaarschot, H. A. Weidenm\"uller, and
  M. R. Zirnbauer, Phys. Rep. {\bf 129} (1985) 367.

\bibitem{bro59}G. E. Brown, Revs. Mod. Phys. {\bf 31} (1959) 893.

\bibitem{lev98}J. Levin, H. Feldman, A. Baer, D. Ben-Hamu, O. Heber,
  D. Zaifman, and Z. Vager (private communication 1998).

\end{references}
\end{document}